\documentclass[aps,pre,twocolumn,groupedaddress,showpacs,superscriptaddress]{revtex4}
\usepackage{amsmath,amssymb}
\usepackage{graphicx,color}

\begin{document}
\title{Derivation of a three-dimensional phase-field-crystal model for liquid crystals from density functional theory}

\author{Raphael Wittkowski, Hartmut L{\"o}wen}
\affiliation{Institut f{\"u}r Theoretische Physik II, Weiche Materie,
Heinrich-Heine-Universit{\"a}t D{\"u}sseldorf, D-40225 D{\"u}sseldorf, Germany}

\author{Helmut R. Brand}
\affiliation{Theoretische Physik III, Universit{\"a}t Bayreuth, D-95540 Bayreuth, Germany}
\date{\today}

\begin{abstract} Using a generalized order parameter gradient expansion 
within density functional theory, we derive a phase-field-crystal
model for liquid crystals composed by apolar particles in three spatial dimensions.
Both the translational density and the orientational direction and ordering are included as order parameters.
Different terms involving gradients in the order parameters in the resulting free energy functional are compared to the macroscopic Ginzburg-Landau approach as well as to the hydrodynamic description for liquid crystals. Our approach provides microscopic expressions for all prefactors in terms of the particle interactions.
Our phase-field-crystal model generalizes the conventional phase-field-crystal model of spherical particles to orientational degrees of freedom and can be used as a starting point to explore phase transitions and interfaces for various liquid-crystalline phases.
\end{abstract}


\maketitle


\section{\label{sec:introduction}Introduction}
The traditional Landau theory of phase transitions in which the free energy is expanded in terms of a convenient order parameter predicts the order and the scaling behavior of bulk phase transitions \cite{Landaubook} in mean field approximation.
The same idea can be used for spatially dependent order parameters
in a free energy functional where gradient expansions lead to a Landau-type description of equilibrium interfaces between two coexisting phases (see e.g.\ \cite{Evans}). This approach was very successful for liquid-gas transition and fluid-fluid phase separation in mixtures and was generalized to nematic and smectic liquid crystalline phases by de Gennes \cite{pgdg71,pgdg73}.
It was also extended to freezing by a multiple \cite{Loewen1,Ohnesorge,Lutsko} or single order parameter theory. The latter involves a gradient expansion up to fourth order in the density field and leads to the so-called phase-field-crystal (PFC) 
model \cite{Mikko,Elder_0,Emmerich}. Depending on the parameter combinations, the PFC model leads to stable periodic density modulations and to constant densities both in two and three spatial dimensions \cite{Jaatinnen_JPCM_2010}. 
Therefore it has been used for large-scale numerical investigations of statics and 
dynamics in the crystalline state including:
the structure and free energy of the fluid crystal interface \cite{Axel,Tapio}, crystal growth dynamics into a supercooled liquid \cite{Laszlo}, the structure \cite{Plapp} and dynamics \cite{Voorhees} of grain boundaries, and  the Asaro-Tiller-Grinfeld instability
\cite{Asaro,Grinfeld,Elder,Voorhees2}. 
The PFC model can be derived from microscopic density functional theory (DFT)
\cite{Evans,Singh:91,Loewen:94,Sven_09,Roth_JPCM_2010} which  describes
crystallization in equilibrium \cite{Ramakrishnan:79,FMT1,White_bear,FMT2}
using a Landau expansion in terms of density modulations \cite{Loewen1,Ohnesorge,Lutsko,Provatas}.
In two spatial dimensions, this derivation was recently generalized to liquid crystalline phase which possess orientational order \cite{Lowen_JPCM_2010} and within quite few parameters a rich topology of the equilibrium phases were found.

In this paper, we generalize the derivation of the phase-field-crystal model for apolar orientational degrees of freedom to three spatial dimensions. We start from microscopic density functional theory and perform a gradient expansion in three order parameters namely the translational density and the orientational direction and ordering (or equivalently the nematic tensor). The result for the static free energy functional is richer than in two dimensions \cite{Lowen_JPCM_2010}. The prefactors of the different gradient terms are expressed as integrals over microscopic correlation functions allowing thus a microscopic determination in terms of the interparticle interactions. 
The results are compared to macroscopic approaches which provide a framework of possible gradient terms allowed from general symmetry principles. 
This phase-field-crystal model generalizes the phase-field-crystal model of spherical particles \cite{Mikko,Elder_0,Emmerich} to orientational degrees of freedom.
It can be used as a starting point to explore phase transitions and interfaces for various liquid-crystalline phases, in particular including plastic and full crystalline phases where the translational density shows a strong ordering.

The paper is organized as follows: in Sec.\ II, we derive the PFC model from  density functional theory by expanding the orientational dependence of the density field up to the first nontrivial order and performing a gradient expansion in the translational coordinate. 
Then, in Sec.\ III, we discuss the relation to symmetry-based approaches. 
We finally discuss possible extensions of the model to more complicated situations and give final conclusions in Sec.\ IV.

\section{\label{sec:derivation}Derivation of the phase-field-crystal model from density functional theory}
Our derivation of the PFC model uses the microscopic static density functional theory for liquid crystals.
We consider $N$ particles with center-of-mass positions 
$\{ \vec{R}_i \,|\, i=1,\dotsc,N \}$ and two orientational degrees of freedom each. 
The actual orientations of these particles are described by a set of unit vectors 
$\{ \hat{u}_i \,|\, i=1,\dotsc,N \}$.
In three dimensions the system has a total volume $V$ and is kept at a finite temperature $T$ in the domain 
$\mathcal{V}\subseteq\mathbb{R}^{3}$.

A pair interaction potential 
$W( \vec{R}_{1} - \vec{R}_{2},{\hat u}_{1}, {\hat u}_{2} )$ between 
two particles 1 and 2 is assumed. We henceforth consider \textit{apolar} and \textit{uniaxial} particles. The apolarity implies the following symmetries:
\begin{equation}
\begin{split}
W(\vec{r},\hat{u}_{1},\hat{u}_{2})  
&= W(-\vec{r},\hat{u}_{1},\hat{u}_{2}) 
= W(\vec{r},-\hat{u}_{1},\hat{u}_{2}) \\
&= W(\vec{r},\hat{u}_{1},-\hat{u}_{2}) 
= W(\vec{r},\hat{u}_{2},\hat{u}_{1}) \;.
\end{split}
\end{equation}
Examples for 
$W(\vec{R}_{1} - \vec{R}_{2},\hat{u}_{1},\hat{u}_{2})$ 
comprise excluded volume interactions 
(e.g.\ hard spherocylinders \cite{Bolhuis,Lowen_HS} or hard ellipsoids \cite{Mulder}), Yukawa segment models \cite{Yukawa_Lowen_1,Yukawa_Lowen_2,Yukawa_Lowen_3} and Gay-Berne potentials \cite{Gay_Berne_1,Gay_Berne_2,Gay_Berne_3}.

The inhomogeneous one-particle density 
\begin{equation}
\rho(\vec{R},\hat{u}) = \left\langle \sum^{N}_{i=1} \delta(\vec{R}-\vec{R}_{i}) 
\delta(\hat{u}-\hat{u}_i) \right\rangle
\end{equation}
provides the joint probability density to find particles at
center-of-mass-position $\vec{R}$ with orientation $\hat{u}$.
For an observable $\mathcal{A}$ the symbol 
\begin{equation}
\begin{split}
\langle\mathcal{A}\rangle = 
&\frac{1}{Z} 
\int_{\mathcal{V}} \mathrm{d}^{3}\vec{R}_{1} \dotsi\!\! \int_{\mathcal{V}} \mathrm{d}^{3}\vec{R}_N
\int_{S_{2}} \mathrm{d}^{2}\hat{u}_{1} \dotsi\!\!  \int_{S_{2}} \mathrm{d}^{2}\hat{u}_{N} 
\, \\ &\times\mathcal{A} 
\exp \! \left[ -\sum_{i\neq j=1}^N 
\frac{W(\vec{R}_i-\vec{R}_j,\hat{u}_i,\hat{u}_j)}{k_{\mathrm{B}}T} \right]
\end{split}
\end{equation}
denotes the normalized canonical average, with the classical 
canonical partition function $Z$, the unit sphere 
$S_{2}=\{\vec{x}\in\mathbb{R}^{3}:\lvert\vec{x}\rvert=1\}$ and 
Boltzmann's constant $k_{\mathrm{B}}$.
Apolar particles involve the symmetry 
$\rho(\vec{R},\hat{u})=\rho(\vec{R},-\hat{u})$.

From classical density functional theory of inhomogeneous fluids we know about the existence of an excess free energy functional $\mathcal{F}_{\mathrm{exc}}$ such that the Landau free energy functional
\begin{equation}
\begin{split}
&\Omega(T,\mathcal{V},\mu,[\rho(\vec{R},\hat{u})]) = 
\mathcal{F}_{\mathrm{id}}(T,\mathcal{V},[\rho(\vec{R},\hat{u})]) \\
&+ \mathcal{F}_{\mathrm{exc}}(T,\mathcal{V},[\rho(\vec{R},\hat{u})]) 
- \int_{\mathcal{V}} \mathrm{d}^{3}\vec{R} \int_{S_{2}} \mathrm{d}^{2}\hat{u}\:\!\mu \rho ({\vec R}, {\hat u})
\end{split}
\end{equation}
is minimal for the equilibrium density field for a given chemical potential $\mu$, temperature $T$ and domain $\mathcal{V}$.
The ideal rotator gas functional $\mathcal{F}_{\mathrm{id}}$ is known exactly:
\begin{equation}
\begin{split}
\mathcal{F}_{\mathrm{id}}(T,\mathcal{V},[\rho(\vec{R},\hat{u})]) = 
&k_{\mathrm{B}}T\int_{\mathcal{V}}\mathrm{d}^{3}\vec{R}\int_{S_{2}}\mathrm{d}^{2}\hat{u}\, \rho(\vec{R},\hat{u}) \\
&\times\Big(\ln\big(\Lambda^{3}\rho(\vec{R},\hat{u})\big)-1\Big) \;. 
\label{eq:Fid} 
\end{split}
\end{equation}
Here, $\Lambda$ is the thermal de Broglie wavelength. 
The excess free energy functional 
$\mathcal{F}_{\mathrm{exc}}(T,\mathcal{V},[\rho(\vec{R},\hat{u})])$  incorporates all correlations and is not known in general, 
but there are several approximations available. 
Well known approximations include the Onsager functional 
\begin{equation}
\begin{split}
&\mathcal{F}_{\mathrm{exc}}(T,\mathcal{V},[\rho(\vec{R},\hat{u})])  
\approx -\frac{k_{\mathrm{B}}T}{2}\int_{\mathcal{V}} \mathrm{d}^{3}\vec{R}_{1} 
\int_{\mathcal{V}} \mathrm{d}^{3}\vec{R}_{2} \\ 
&\qquad\times  
\int_{S_{2}} \mathrm{d}^{2}\hat{u}_{1}  
\int_{S_{2}} \mathrm{d}^{2}\hat{u}_{2} \,\rho(\vec{R}_{1},\hat{u}_{1})\rho(\vec{R}_{2},\hat{u}_{2}) \\
&\qquad\times\left(\exp \! \left[-\frac{W(\vec{R}_{1}-\vec{R}_{2},\hat{u}_{1},\hat{u}_{2})}{k_{\mathrm{B}}T}\right] -1\right) 
\end{split}
\end{equation}
that becomes asymptotically exact in the low density limit \cite{Frenkel}, 
the mean-field approximation
\begin{equation}
\begin{split}
\mathcal{F}_{\mathrm{exc}}(T,\mathcal{V},[\rho(\vec{R},\hat{u})]) \approx \frac{1}{2}\int_{\mathcal{V}} \mathrm{d}^{3}\vec{R}_{1} \int_{\mathcal{V}} \mathrm{d}^{3}\vec{R}_{2} 
\int_{S_{2}} \mathrm{d}^{2}\hat{u}_{1}& \\
\times\int_{S_{2}} \mathrm{d}^{2}\hat{u}_{2} \, 
W(\vec{R}_{1}-\vec{R}_{2},\hat{u}_{1},\hat{u}_{2}) 
\rho(\vec{R}_{1},\hat{u}_{1})\rho(\vec{R}_{2},\hat{u}_{2})& 
\end{split}
\end{equation}  
which becomes asymptotically exact at high densities for bounded potentials \cite{Rex_08} and the 
\textit{Ramakrishnan-Yussouff approximation} \cite{Ramakrishnan:79}  
\begin{equation}
\begin{split}
&\mathcal{F}_{\mathrm{exc}}(T,\mathcal{V},[\rho(\vec{R},\hat{u})]) \approx 
- \frac{k_{\mathrm{B}}T}{2} \int_{\mathcal{V}}\mathrm{d}^{3}\vec{R}_{1} \int_{\mathcal{V}}\mathrm{d}^{3}\vec{R}_{2} \\ &\qquad\times\int_{S_{2}}\mathrm{d}^{2}\hat{u}_{1}\int_{S_{2}}\mathrm{d}^{2}\hat{u}_{2} \, 
c^{(2)}(\vec{R}_{1}-\vec{R}_{2},\hat{u}_{1},\hat{u}_{2}) \\
&\qquad\times\Big(\rho(\vec{R}_{1},\hat{u}_{1})-\overline{\rho}\Big)
\Big(\rho(\vec{R}_{2},\hat{u}_{2})-\overline{\rho}\Big) \,. 
\end{split}
\label{eq:RY}
\end{equation}
This approximation is used in the following. 
It can be viewed as a truncated density expansion in the density 
difference $\rho(\vec{R},\hat{u})-\overline{\rho}$ around a mean density $\overline{\rho}$ with the direct correlation function $c^{(2)}$ of a reference fluid. 
More accurate forms for $\mathcal{F}_{\mathrm{exc}}$ for hard particles are given by weighted-density-approximations \cite{Holyst,Graf} or follow from the fundamental-measure theory \cite{FMT2}.
As a further approximation we only consider weak anisotropies in the orientations. Thus, the leading terms in the density parametrization for uniaxial particles are
\begin{equation}
\rho(\vec{R},\hat{u}) \approx \overline{\rho}
\Big(1 + \psi_{1}(\vec{R}) + \psi_{2}(\vec{R}) \operatorname{P}_{2} 
\!\big(\hat{u}_{0}(\vec{R})\cdot\hat{u}\big)\Big)
\label{eq:XXX}
\end{equation}
with $\operatorname{P}_{2}(x)=\frac{1}{2}(3x^{2}-1)$ denoting the second Legendre polynomial.
In this expression, the real-valued dimensionless orientationally averaged density is represented by
\begin{equation}
\psi_{1}(\vec{R}) = \frac{1}{4\pi\overline{\rho}}\int_{S_{2}}\mathrm{d}^{2}\hat{u} \big(\rho(\vec{R},\hat{u})-\overline{\rho}\big) \;. 
\label{eq:psi1}
\end{equation}
It is identical to the original treatment of the PFC model \cite{Mikko,Elder_0}. The dimensionless field
\begin{equation}
\psi_{2}(\vec{R}) = \frac{5}{4\pi\overline{\rho}}\int_{S_{2}}\mathrm{d}^{2}\hat{u} \:\! 
\rho(\vec{R},\hat{u}) \operatorname{P}_{2}\!\big(\hat{u}_{0}(\vec{R})\cdot\hat{u}\big) \;,  
\label{eq:psi2}
\end{equation}
on the other hand, measures the local degree of orientational order. For apolar particles, the leading anisotropic contribution is the third term in the big brackets on the right-hand-side of Eq.\ \eqref{eq:XXX}.
Furthermore, the unit vector field $\hat{u}_0(\vec{R})$ defines the local director of the orientation field \cite{footnote_Referee_B}.

To begin with the derivation of the static free energy functional 
$\mathcal{F}=\mathcal{F}_{\mathrm{id}}+\mathcal{F}_{\mathrm{exc}}$ 
we insert the truncated expansion \eqref{eq:XXX} into Eq.\ \eqref{eq:Fid} and expand the logarithm up to third order. 
After performing the angular integration the approximation  
\begin{equation}
\begin{aligned}
\mathcal{F}_{\mathrm{id}}[\psi_{1},\psi_{2},\hat{u}_{0}] \approx F_{0} + k_{\mathrm{B}}T\overline{\rho}\,2\pi\int_{\mathcal{V}}\mathrm{d}^{3}\vec{R}\quad& \\ 
\times\bigg( 2\psi_{1} + \psi^{2}_{1} - \frac{\psi^{3}_{1}}{3} + \frac{\psi^{4}_{1}}{6} + \frac{\psi^{2}_{2}}{5} - \frac{\psi_{1}\psi^{2}_{2}}{5}& \\ 
+ \frac{\psi^{2}_{1}\psi^{2}_{2}}{5} - \frac{2\psi^{3}_{2}}{105} + \frac{4\psi_{1}\psi^{3}_{2}}{105} + \frac{\psi^{4}_{2}}{70}&\bigg) 
\end{aligned}
\label{eq:Fidx}
\end{equation}
with
\begin{equation}
F_{0} = k_{\mathrm{B}}T\overline{\rho}\,4\pi V\big(\ln(\Lambda^{3}\overline{\rho})-1\big) 
\quad\text{and}\quad
V=\int_{\mathcal{V}}\mathrm{d}^{3}\vec{R} 
\end{equation}
is obtained. 
Next, we derive the excess free energy functional. 
For this purpose we insert Eq.\ \eqref{eq:XXX} into the Ramakrishnan-Yussouff approximation \eqref{eq:RY} and decompose the direct correlation function into spherical harmonics $\operatorname{Y}_{l,m}(\hat{u})$ up to second order:
\begin{equation}
\begin{split}
c^{(2)}(\vec{R},\hat{u}_{1},\hat{u}_{2}) \approx 
&\sum^{2}_{\begin{subarray}{c}l_{j}=0\\1\leqslant j\leqslant 2\end{subarray}}\sum^{l_{j}}_{\begin{subarray}{c}m_{j}=-l_{j}\\1\leqslant j\leqslant 2\end{subarray}} 
\tilde{c}^{(2)}_{l_{1},l_{2},m_{1},m_{2}}(\vec{R}) \\ 
&\times\operatorname{Y}_{l_{1},m_{1}}(\hat{u}_{1}) 
\operatorname{Y}_{l_{2},m_{2}}(\hat{u}_{2}) \;. 
\end{split}
\label{eq:c2}
\end{equation}
The angular integration leads to the final expression 
\begin{equation}
\begin{aligned}
&\mathcal{F}_{\mathrm{exc}}[\psi_{1},\psi_{2},\hat{u}_{0}]
\approx - k_{\mathrm{B}}T \overline{\rho}^{2} 8\pi^{2} \int_{\mathcal{V}}\mathrm{d}^{3}\vec{R}_{1}\int_{\mathcal{V}}\mathrm{d}^{3}\vec{R}_{2} \\ 
&\quad\times\sum^{1}_{\begin{subarray}{c}l_{j}=0\\1\leqslant j\leqslant 2\end{subarray}}
\sum^{2l_{j}}_{\begin{subarray}{c}m_{j}=-2l_{j}\\1\leqslant j\leqslant 2\end{subarray}}  5^{-l_{1}-l_{2}}\,\tilde{c}^{(2)}_{2l_{1},2l_{2},m_{1},m_{2}}(\vec{R}_{1}-\vec{R}_{2})  \\ 
&\quad\times
\operatorname{Y}_{2l_{1},m_{1}}\big(\hat{u}_{0}(\vec{R}_{1})\big) 
\:\!\psi_{l_{1}+1}(\vec{R}_{1}) \\ 
&\quad\times\operatorname{Y}_{2l_{2},m_{2}}\big(\hat{u}_{0}(\vec{R}_{2})\big) 
\:\!\psi_{l_{2}+1}(\vec{R}_{2}) \;. \\ 
\end{aligned} 
\label{eq:Fexc}
\end{equation}
Here, the expansion coefficients 
\begin{equation}
\begin{split}
\tilde{c}^{(2)}_{l_{1},l_{2},m_{1},m_{2}}(\vec{R}) = 
&\int_{S_{2}}\mathrm{d}^{2}\hat{u}_{1} 
\int_{S_{2}}\mathrm{d}^{2}\hat{u}_{2} \, c^{(2)}\big(\vec{R},\hat{u}_{1},\hat{u}_{2}\big) \\
&\times\overline{\operatorname{Y}}_{l_{1},m_{1}}\big(\hat{u}_{1}\big) 
\overline{\operatorname{Y}}_{l_{2},m_{2}}\big(\hat{u}_{2}\big) 
\end{split}
\label{eq:cw} 
\end{equation}
are not independent. If they are further decomposed into a series of spherical harmonics with respect to the remaining orientational unit vector 
$(\vec{R}_{1}-\vec{R}_{2})/\lvert\vec{R}_{1}-\vec{R}_{2}\rvert$, it is possible to use the identities 
\begin{equation}
\begin{aligned}
&c^{(2)}(R\hat{u},\hat{u}_{1},\hat{u}_{2}) = 
\sum^{\infty}_{l_{1},l_{2},l=0} 
\omega_{l_{1},l_{2},l}(R) 
\sum^{l_{j}}_{\begin{subarray}{c}m_{j}=-l_{j}\\1\leqslant j\leqslant 2\end{subarray}} 
\sum^{l}_{m=-l}  \\
&\qquad\times C(l_{1},l_{2},l,m_{1},m_{2},m) \operatorname{Y}_{l_{1},m_{1}}(\hat{u}_{1}) \\
&\qquad\times \operatorname{Y}_{l_{2},m_{2}}(\hat{u}_{2}) \, 
\overline{\operatorname{Y}}_{l,m}(\hat{u}) \\
\end{aligned} 
\label{eq:CG_Entwicklung}
\end{equation}
and 
\begin{equation}
\begin{aligned}
&\omega_{l_{1},l_{2},l}(R) 
= \sqrt{\frac{4\pi}{2l+1}} \int_{S_{2}}\! \mathrm{d}^{2}\hat{u}_{1} \int_{S_{2}}\! \mathrm{d}^{2}\hat{u}_{2} \!\!\!\!\!\!\!\! \sum^{\min\{l_{1},l_{2}\}}_{m=-\min\{l_{1},l_{2}\}} \\
&\qquad\times C(l_{1},l_{2},l,m,-m,0) 
\,\overline{\operatorname{Y}}_{l_{1},m}(\hat{u}_{1}) 
\,\overline{\operatorname{Y}}_{l_{2},-m}(\hat{u}_{2}) \\
&\qquad\times c^{(2)}(R\hat{e}_{z},\hat{u}_{1},\hat{u}_{2}) \\
\end{aligned} 
\label{eq:CG_omega}
\end{equation}
to derive necessary relations between the expansion coefficients from the properties of the Clebsch-Gordan coefficients $C(l_{1},l_{2},l,m_{1},m_{2},m)$ \cite{GrafLoewen1998}. 
Now a gradient expansion is performed \cite{Provatas} up to fourth order in the $\psi_{1}\psi_{1}$ terms of Eq.\ \eqref{eq:Fexc} and up to second order in the $\psi_{1}\psi_{2}$ and $\psi_{2}\psi_{2}$ terms. We assume that the highest gradient terms ensure stability.
By partial integration and omission of surface terms one obtains the result 
\begin{equation}
\begin{aligned}
&\mathcal{F}_{\textrm{exc}}[\psi_{1},\psi_{2},\hat{u}_{0}] \approx 
\frac{1}{2} \int_{\mathbb{R}^{3}} \mathrm{d}^{3}\vec{R} \\
&\;\,\times\biggl( 
A_{1}\psi^{2}_{1} + A_{2}\big(\vec{\nabla}\psi_{1}\big)^{2} 
+ A_{3}\big(\triangle\psi_{1}\big)^{2} \\ 
&\qquad+ B_{1}\psi^{2}_{2} 
+ \frac{1}{9}\big(\tilde{K}_{1}+2\tilde{K}_{2}\big) \big(\vec{\nabla}\psi_{2}\big)^{2} \\ &\qquad+ \frac{1}{3}\big(\tilde{K}_{1}-\tilde{K}_{2}\big) \big( \hat{u}_{0}\!\!\:\cdot\!\!\:\vec{\nabla}\psi_{2} \big)^{2} \\ 
&\qquad+ B_{2}\Big( 
\vec{\nabla}\psi_{1}\!\!\:\cdot\!\!\:\vec{\nabla}\psi_{2} 
-3 \big(\hat{u}_{0}\!\!\:\cdot\!\!\:\vec{\nabla}\psi_{1}\big) 
\big(\hat{u}_{0}\!\!\:\cdot\!\!\:\vec{\nabla}\psi_{2}\big) \\  &\qquad\qquad-3\psi_{2}\vec{\nabla}\psi_{1} \!\!\:\cdot\!\!\: 
\big(\big(\hat{u}_{0}\!\!\:\cdot\!\!\:\vec{\nabla}\big)\hat{u}_{0} 
+ \hat{u}_{0}\big(\vec{\nabla} \!\!\:\cdot\!\!\: \hat{u}_{0}\big)\big)\Big) \\ 
&\qquad+ 2\psi_{2}\vec{\nabla}\psi_{2} \!\!\:\cdot\!\!\: 
\Big( \big(\tilde{K}_{2}-\tilde{K}_{1}\big) 
\big(\big(\hat{u}_{0}\!\!\:\cdot\!\!\:\vec{\nabla}\big)\hat{u}_{0}\big) \\
&\qquad\qquad+ \frac{1}{3}\big(2\tilde{K}_{1}+\tilde{K}_{2}\big) 
\big(\hat{u}_{0}\big(\vec{\nabla}\!\!\:\cdot\!\!\:\hat{u}_{0}\big)\big) 
\Big) \\ 
&\qquad+ \psi^{2}_{2} 
\Big( \tilde{K}_{1} 
\big(\vec{\nabla}\!\!\:\cdot\!\!\:\hat{u}_{0}\big)^{2} + \tilde{K}_{2} 
\big(\hat{u}_{0}\!\!\:\cdot\!\!\:\big(\vec{\nabla}\!\times\!\hat{u}_{0}\big)\big)^{2} \\ 
&\qquad\qquad+ \tilde{K}_{1} 
\big(\hat{u}_{0}\!\times\!\big(\vec{\nabla}\!\times\!\hat{u}_{0}\big)\big)^{2}
\Big)\biggr) 
\end{aligned}
\label{eq:Fexc_Glg}
\end{equation}
for $\mathcal{V}=\mathbb{R}^{3}$. 
The coefficients are given by
\begin{equation}%
\allowdisplaybreaks\begin{split}%
A_{1} &= -k_{\mathrm{B}}T\overline{\rho}^{2}4\pi\mathrm{M}^{(0)}_{0,0} \;,\\
A_{2} &= k_{\mathrm{B}}T\overline{\rho}^{2} \frac{2\pi}{3}\mathrm{M}^{(2)}_{0,0} \;,\\
A_{3} &= -k_{\mathrm{B}}T\overline{\rho}^{2} \frac{\pi}{30}\mathrm{M}^{(4)}_{0,0} \;,\\
B_{1} &= -k_{\mathrm{B}}T\overline{\rho}^{2} \frac{4\pi}{25} \Big(\mathrm{M}^{(0)}_{2,0}-2\mathrm{M}^{(0)}_{2,1}+2\mathrm{M}^{(0)}_{2,2}\Big) \;,\\
B_{2} &= -k_{\mathrm{B}}T\overline{\rho}^{2} \frac{4\pi}{15\sqrt{5}}\mathrm{M}^{(2)}_{0,2} \;,\\
\tilde{K}_{1} &= k_{\mathrm{B}}T\overline{\rho}^{2} \frac{2\pi}{175} \Big(9\mathrm{M}^{(2)}_{2,0}-16\mathrm{M}^{(2)}_{2,1}+10\mathrm{M}^{(2)}_{2,2}\Big) \;,\\
\tilde{K}_{2} &= k_{\mathrm{B}}T\overline{\rho}^{2} \frac{2\pi}{175} \Big(3\mathrm{M}^{(2)}_{2,0}-10\mathrm{M}^{(2)}_{2,1}+22\mathrm{M}^{(2)}_{2,2}\Big)  
\end{split}%
\label{subeq:Fexc_Koeffizienten}%
\end{equation}%
and depend on the moments  
\begin{equation}%
\allowdisplaybreaks
\begin{split}%
\mathrm{M}^{(n)}_{0,0} &= 4\pi \int^{\infty}_{0} \mathrm{d}R R^{n+2} \tilde{c}^{(2)}_{0,0,0,0}(R) \;,\\
\mathrm{M}^{(n)}_{0,2} &= 4\pi \int^{\infty}_{0} \mathrm{d}R R^{n+2} \tilde{c}^{(2)}_{0,2,0,0}(R\hat{e}_{z}) \;,\\
\mathrm{M}^{(n)}_{2,0} &= 4\pi \int^{\infty}_{0} \mathrm{d}R R^{n+2} \tilde{c}^{(2)}_{2,2,0,0}(R\hat{e}_{z}) \;,\\
\mathrm{M}^{(n)}_{2,1} &= 4\pi \int^{\infty}_{0} \mathrm{d}R R^{n+2} \tilde{c}^{(2)}_{2,2,1,-1}(R\hat{e}_{z}) \;,\\
\mathrm{M}^{(n)}_{2,2} &= 4\pi \int^{\infty}_{0} \mathrm{d}R R^{n+2} \tilde{c}^{(2)}_{2,2,2,-2}(R\hat{e}_{z}) 
\end{split}%
\label{eq:Fexc_Momente}%
\end{equation}%
of the expansion coefficients of the direct correlation function, where $\hat{e}_{z}$ is the cartesian unit vector codirectional with the $z$-axis. 
These moments in turn depend on the thermodynamic conditions expressed by $T$ and $\overline{\rho}$.

The functional \eqref{eq:Fexc_Glg} contains a few simpler models as special cases which follow by setting $\psi_{1}$, $\psi_{2}$ and $\hat{u}_{0}$ successively to zero or to a constant. 
In detail, if $\psi_{1}=0$ and $\psi_{2}$ and $\hat{u}_{0}$ are constant, the PFC model corresponds to a completely isotropic system and all terms in Eq.\ \eqref{eq:Fexc_Glg} vanish or are constant ($\sim B_{1}$). 
Alternatively, when $\psi_{1}=0$, $\psi_{2}$ is constant and $\hat{u}_{0}$ is space-dependent, the Frank free energy functional \cite{pgdgbook} is recovered. 
Up to the common prefactor $\psi^{2}_{2}$ the Frank constants for splay, twist and bend are $\tilde{K}_{1}$, $\tilde{K}_{2}$ and $\tilde{K}_{3}=\tilde{K}_{1}$, respectively.
The more general case with $\psi_{1}=0$ and space-dependent $\psi_{2}$ and $\hat{u}_{0}$ can be referred to as constant-density approximation. 
If then $\hat{u}_{0}$ is a constant unit vector, the constant-density excess free energy functional reduces to a simple gradient expansion of second order for $\psi_{2}$, where only the terms four to six are not vanishing. 
Otherwise, for a space-dependent $\hat{u}_{0}$, the PFC model corresponds to the Landau-de Gennes free energy \cite{pgdgbook} for uniaxial nematics. 
More complex models are recovered for non-constant densities, i.e.\ for $\psi_{1}=\psi_{1}(\vec{R})$. 
If only $\psi_{1}$ is space-dependent, the PFC model has no orientational dependence and is equivalent to the three-dimensional extension of the PFC-model of K.\ R.\ Elder and coworkers \cite{Provatas} for isotropic particles. 
With a space-dependent $\hat{u}_{0}$, this model is extended to the free energy for uniaxial nematics with a constant amount of ordering. 
Also for non-constant scalar fields $\psi_{1}$ and $\psi_{2}$ but a constant nematic director $\hat{u}_{0}$, the model is much simpler than the full excess free energy functional in Eq.\ \eqref{eq:Fexc_Glg}, because the computationally expensive terms that describe the couplings of $\vec{\nabla}\psi_{1}$ and $\vec{\nabla}\psi_{2}$ with derivatives of the nematic director as well as the very involved Frank free energy drop out. 

In the full excess free energy functional \eqref{eq:Fexc_Glg} all these special cases are properly comprised. 
This new functional clarifies the relation between already existing simpler PFC models, contains the appropriate couplings of the fields $\psi_{1}$, $\psi_{2}$ and $\hat{u}_{0}$, relates the constant prefactors of the terms in Eq.\ \eqref{eq:Fexc_Glg} to the direct correlation function and is therefore the main result of this paper.

\section{Relation to symmetry-based approaches}
\label{sec:diagram}
In this section we make contact between the three-dimensional PFC model for liquid crystals based on density functional theory with two macroscopic symmetry-based approaches, namely the Ginzburg-Landau description and generalized hydrodynamics.
The goal is to compare the central results of this paper summarized in Eqs.\ \eqref{eq:Fexc_Glg} to \eqref{eq:Fexc_Momente} with corresponding results from Ginzburg-Landau analysis appropriate as a mean field description in the vicinity of phase transitions and the hydrodynamic description applicable for long wavelengths (continuum approximation) and low frequencies.

For the contributions associated with density variations
and their gradients given in the second line of Eq.\ \eqref{eq:Fexc_Glg}
this can be done very easily.
From the Ginzburg-Landau description for  the smectic A - nematic
description we have for the corresponding terms \cite{mpb1}
in the energy density
\begin{equation}
\label{Aiso}
  \frac{1}{2} \alpha   |\psi|^{2}
+ \frac{1}{2} b_{1}    |\vec{\nabla}_{i}\psi|^{2}
+ \frac{1}{2} b_{2}    |\triangle\psi|^{2}
\end{equation}
where we have used the smectic order parameter $\psi=\psi_{0}\exp(-i \phi)$ with magnitude $\psi_{0}$ and phase $\phi$ introduced by de Gennes \cite{pgdg73,pgdgbook}, which is directly proportional to density variations (compare, for example \cite{cl}). 
Comparing the second line of Eq.\ \eqref{eq:Fexc_Glg} and Eq.\ \eqref{Aiso}, we can identify $A_{1}$, $A_{2}$ and $A_{3}$ with $\alpha$, $b_{1}$ and $b_{2}$, respectively.

In addition, we can also make contact with the bulk description of
smectic A, where one uses the layer displacement $u$ parallel to the
layer normal as a hydrodynamic variable
\cite{pgdg69,pgdgbook}. For the gradient terms associated 
with the layer displacement, $u$, which is proportional to phase changes,
one has in the energy density 
\begin{equation}
  \frac{1}{2} B (\vec{\nabla}_{\parallel} u)^{2} 
+ \frac{1}{2} K (\vec{\nabla}^{2}_{\perp} u)^{2} 
\label{Abulk} 
\end{equation}
where in Eq.\ \eqref{Abulk} the contribution $\sim B$ is associated with the 
compressibility of the layering and the contribution $\sim K$ is connected with bending of the layering. 
Since the macroscopic description is dealing with the bulk of 
the smectic A phase, the uniaxial anisotropy is reflected in
the terms given in Eq.\ \eqref{Abulk}. 

For the terms associated exclusively with orientational order in Eq.\ \eqref{eq:Fexc_Glg} in lines 3, 4 and 7-10 we start the comparison with the continuum description of the bulk phase for which we have for the analogous terms in the energy density
\begin{equation}
\begin{split}
K_{1} (\vec{\nabla} \cdot \vec{n})^{2} &+ 
K_{2} (\vec n \cdot [\vec{\nabla} \times \vec n ])^{2} + 
K_{3} (\vec n \times [\vec{\nabla} \times \vec n ])^{2} \\ 
&+ L_{\parallel} (n_i \vec{\nabla}_i S)^{2}
+ L_{\perp} \delta_{ij}^{\mathrm{tr}}(\vec{\nabla}_i S) (\vec{\nabla}_j S) \\
&+ M (\vec{\nabla}_i S) 
[ \delta_{ik}^{\mathrm{tr}} n_j + \delta_{jk}^{\mathrm{tr}} n_i ] (\vec{\nabla}_j n_k) \;. 
\end{split}
\label{nemgrad}
\end{equation}
In Eq.\ \eqref{nemgrad} the first line is connected to gradients of the director field, $\vec n$. It contains splay, twist and bend
and goes back to Frank's pioneering paper \cite{frank,pgdgbook}.
Lines 2 and 3 are associated with gradients of the nematic modulus, $S$, and with a coupling term, $\sim M$, between gradients of the director and gradients of the modulus \cite{bk86,bp87}.

The Frank free energy can be easily compared with lines 9 and 10 of Eq.\ \eqref{eq:Fexc_Glg}. We identify $\psi_{2}$ in the last section with the nematic modulus, $S$, and $\hat{u}_{0}$ with the director field, $\vec{n}$.
The splay constant $K_{1}$ in Eq.\ \eqref{nemgrad} reads $2\tilde{K}_{1}\psi^{2}_{2}$, the twist elastic constant in Eq.\ \eqref{nemgrad} reads $2\tilde{K}_{2}\psi^{2}_{2}$ and the bend elastic constant in Eq.\ \eqref{nemgrad} is $2\tilde{K}_{1}\psi^{2}_{2}$.
A similar comparison can be performed for the expressions given in lines 
3, 4, 7 and 8 of Eq.\ \eqref{eq:Fexc_Glg}. We obtain for the contribution $\sim M$ in Eq.\ \eqref{nemgrad} $\psi_{2}(\tilde{K}_{2}-\tilde{K}_{1})$ and for the gradient terms of the nematic modulus $L_{\parallel}$ and $L_{\perp}$ are given by $L_{\parallel} = \frac{2}{9}(4\tilde{K}_{1} - \tilde{K}_{2})$ and $L_{\perp} = \frac{2}{9}(\tilde{K}_{1} +  2 \tilde{K}_{2})$. 

We thus arrive at the conclusion that instead of 6 independent coefficients in the symmetry-based continuum description only two independent ones are left over in the PFC model.
This reduction in the number of independent coefficients is a well known feature of approximate approaches. 
It is even known in fields such as superfluid $^{3}$He where one finds a reduction from 6 to 3 coefficients for analogous terms \cite{bk86,mcc75}.
In this connection it turns out to be quite instructive to compare the terms associated with gradients of the orientational order in Eq.\ \eqref{eq:Fexc_Glg} with the gradient terms in the Ginzburg-Landau description of
the nematic - isotropic phase transition given by de Gennes \cite{pgdg71} with the same terms also occuring in the Ginzburg-Landau description of the smectic A - isotropic phase transition.

In the vicinity of the isotropic - uniaxial nematic phase transition one takes traditionally the second order traceless tensor $Q_{ij}$ \cite{pgdg71} as the order parameter. It vanishes in equilibrium, $< Q_{ij} > \equiv 0$ in the isotropic phase and assumes in the uniaxial nematic phase the structure \cite{pgdgbook}
$ Q_{ij} = S (n_{i}n_{j} -\frac{1}{d} \delta_{ij})$, with $d=2$ in two dimensions and $d=3$ in three dimensions.
The gradients in $Q_{ij}$ give rise to the gradient energy
\begin{equation}
F_{Q} = \int_{\mathbb{R}^{3}} \mathrm{d}^{3}\vec{R}\, 
L_{ijklmn} (\vec{\nabla}_i Q_{jk})(\vec{\nabla}_l Q_{mn}) \;. 
\label{Q}
\end{equation}
In the uniaxial nematic phase the tensor $L_{ijklmn}$ has six independent coefficients. Using the decomposition, for example, in three dimensions 
$Q_{ij} = S (n_{i} n_{j} -\frac{1}{3} \delta_{ij})$ one obtains in total six
coefficients as above: 3 Frank coefficients for the pure deformations
of the director field, 2 coefficients for the deformations 
of the modulus $S$, and one coupling term between gradients of the director and gradients of the modulus. 
In the isotropic phase this expression reduces to \cite{pgdg71}
\begin{equation}
\begin{split}
F_{Q_{\mathrm{iso}}} = \int_{\mathbb{R}^{3}} \mathrm{d}^{3}\vec{R}\, 
&\bigl[L_{1} (\vec{\nabla}_i Q_{jk})(\vec{\nabla}_i Q_{jk}) \\
&+ L_{2} (\vec{\nabla}_i Q_{ik})(\vec{\nabla}_j Q_{jk})\bigr]
\end{split}
\label{Qiso}
\end{equation}
and thus to the same number of independent coefficients as in the PFC approach given above. 
We close the discussion of the terms associated purely with orientational order by pointing out that the first term on the third line in Eq.\ \eqref{eq:Fexc_Glg} is the analog of the term $\sim\beta_{1}$ in Eq.\ (2a) of Ref.\ \cite{bp87}. 

Next we compare the results given in Eq.\ \eqref{eq:Fexc_Glg} for the coupling terms between gradients of the density and gradients of the orientational order with the results of the two macroscopic symmetry-based approaches. 
These terms are listed in lines 5 and 6 of Eq.\ \eqref{eq:Fexc_Glg} and are proportional to $B_{2}$. 
For spatial gradients in the director field coupling to spatial variations
in the density we find \cite{pb80,bp87}
\begin{equation}
F_{\mathrm{nc}} = \int_{\mathbb{R}^{3}} \mathrm{d}^{3}\vec{R}\, \lambda^{\rho} (\vec{\nabla}_i \rho)
[ \delta_{ik}^{\mathrm{tr}} n_j + \delta_{jk}^{\mathrm{tr}} n_i \bigr] (\vec{\nabla}_j n_k) 
\label{ncoup}
\end{equation}
where the transverse Kronecker delta projects onto the plane 
perpendicular to the preferred direction $\vec{n}$:
$\delta_{ij}^{\mathrm{tr}} = \delta_{ij} - n_i n_j$.
By comparison with Eq.\ \eqref{eq:Fexc_Glg} we find $\lambda^{\rho} = - 3 B_{2} \psi_{2}$.
Finally we have for the terms coupling gradients of the order parameter modulus to gradients of the density \cite{bp87}
\begin{equation}
F_{\mathrm{Sc}} = \int_{\mathbb{R}^{3}} \mathrm{d}^{3}\vec{R}\, 
N_{ij}^{\rho} (\vec{\nabla}_i S) (\vec{\nabla}_j \rho) 
\label{Scoup}
\end{equation}
where the second rank tensor $N^{\rho}$ is of the standard uniaxial form
$N^{\rho}_{ij} = N_{1}^{\rho} n_i n_j + N_{2}^{\rho} \delta_{ij}^{\mathrm{tr}}$.
A comparison with Eq.\ \eqref{eq:Fexc_Glg} yields $N_{1} = -2 B_{2}$ and $N_{2} = B_{2}$. 
The coupling terms listed in Eqs.\ \eqref{ncoup} and \eqref{Scoup} exist in both, two and three spatial dimensions. 
Thus in comparison to the hydrodynamic description of the bulk behavior, which is characterized by three independent coefficients, we find one independent coefficient in the PFC model. 
In the framework of a Ginzburg-Landau approach using the orientational
order parameter $Q_{ij}$ we find in the isotropic phase
\begin{equation}
F_{Q_{\mathrm{rhoiso}}} = \int_{\mathbb{R}^{3}} \mathrm{d}^{3}\vec{R}\, 
P^{\xi}(\vec{\nabla}_i Q_{jk})(\vec{\nabla}_l \rho) 
(\delta_{ij} \delta_{kl} + \delta_{ik} \delta_{jl}) 
\label{Qrhoiso}
\end{equation}
and thus one independent coefficient - as has also been the case for the PFC model.

In conclusion we find that the PFC model expanding the generalized energy up to quadratic order in orientational and density variations and their gradients can be compared easily with corresponding terms in
the symmetry-based Ginzburg-Landau description in the isotropic phase.
It also emerges that the number of independent coefficients obtained in the framework of a PFC model based on a density functional approach typically contains a smaller number of independent coefficients than predicted from generalized hydrodynamics.
This feature is shared by other approximate approaches such as BCS for superfluid $^{3}$He, but has the advantage to predict concrete values for these coefficients, which are left as unknown parameters in a hydrodynamic description.

\section{\label{sec:conclusions}Conclusions and possible extensions}
In conclusion, we derived a phase-field-crystal model for liquid crystals in three dimensions from density functional theory. Two approximations are involved: first the density functional is approximated by a truncated functional Taylor expansion similar in spirit to the Ramakrishnan-Yussouff theory. Then a generalized gradient expansion in the order parameters is performed which leads to a local free energy functional. 
There are three order parameters, namely the translational density which 
corresponds to the scalar phase-field variable $\psi_{1}$ in the traditional phase-field-crystal model, an orientational direction given by a three-dimensional unit vector $\hat u_{0}$ and the nematic order parameter 
$\psi_{2}$. In the two latter quantities the gradient expansion is performed up to second order while it is done to fourth order in $\psi_{1}$.
This ensures that the traditional phase-field-crystal model \cite{Mikko,Elder_0} is recovered as a special case in which there is no orientational dependence of the full density. The additional terms are all in accordance with macroscopic approaches based on symmetry considerations \cite{pb95,bp87}. The full static free energy functional allows for a wealth of stable liquid crystalline phases.
How the phase diagram depends on the prefactors should be explored by further numerical studies.
Once the stable phases are known, the structure of interfaces between two coexisting phases can be addressed, not only the isotropic-nematic interface \cite{Schmid,Beek,Harnau} but also interfaces which involve one or two translational ordered crystalline phases.

The analysis presented here can be generalized or extended to quite a number of different situations.
First of all one could in principle include higher order gradients both in $\psi_{1}$ as well as in $\hat{u}_{0}$ and $\psi_{2}$. 
The former is in particularly mandatory if a more realistic description of the translational crystalline density field is targeted which is highly peaked in a three-dimensional solid \cite{EPL_1993_Ohnesorge} or if higher-order orientational distributions should be resolved which is relevant for smectic A phases \cite{vanRoij}.

Second, the generalization to dynamics is in principle straightforward following the lines given in two spatial dimensions in Ref.\ \cite{Lowen_JPCM_2010}. For Brownian dynamics, the dynamical density functional theory \cite{Marconi,Archer_Evans,Pep} was generalized to orientational dynamics \cite{Rex_08} and can be used as a starting point to derive the order parameter dynamics. 
In general, the dynamics for $\psi_{1}$ is conserved while that for 
$\hat u_{0}$ and $\psi_{2}$ is not. However, though the derivation can be done in principle, it turns out in practice that the actual equations of motions for the order parameters involve a huge number of terms such that it is too tedious to state them all explicitly.
The dynamical extension could in principle be applied to the dynamics of topological defects \cite{Muthukumar} and to interfacial dynamics near three-phase coexistence \cite{Bechhoefer}. Also, the dynamics in orientational glasses \cite{Renner_Barrat} could be explored.

Third, polar particles with a head-tail asymmetry will violate the 
symmetry conditions used here for the director field. There should not be a principle obstacle to derive the free energy functional for polar particles as well.
Furthermore, throughout the paper, we assumed uniaxiality. Biaxiality in the orientational distribution could also be included in the gradient expansions at the expense of more coefficients entering the picture.
It will also be most interesting to see how the treatment of 
bond-orientational order in the framework of the PFC-model compares with
the results available from continuum-type approaches.

Finally it would be interesting to generalize the analysis to active particles which are driven by a constant propagation speed along their orientation \cite{Ramaswamy,Markus_Baer}. A dynamical density functional 
approach was recently \cite{Wensink_08} proposed for active particles which could be used as a microscopic starting point to derive systematically gradient expansions.

\begin{acknowledgments}
We thank C. V. Achim, S. van Teeffelen, H. Emmerich and U. Zimmermann for helpful discussions. This work has been supported by the DFG through the DFG priority program SPP 1296 and by SFB TR6 (project D3).
\end{acknowledgments}

\bibliography{References}

\providecommand*{\mcitethebibliography}{\thebibliography}
\csname @ifundefined\endcsname{endmcitethebibliography}
{\let\endmcitethebibliography\endthebibliography}{}
\begin{mcitethebibliography}{69}
\providecommand*{\natexlab}[1]{#1}
\providecommand*{\mciteSetBstSublistMode}[1]{}
\providecommand*{\mciteSetBstMaxWidthForm}[2]{}
\providecommand*{\mciteBstWouldAddEndPuncttrue}
  {\def\EndOfBibitem{\unskip.}}
\providecommand*{\mciteBstWouldAddEndPunctfalse}
  {\let\EndOfBibitem\relax}
\providecommand*{\mciteSetBstMidEndSepPunct}[3]{}
\providecommand*{\mciteSetBstSublistLabelBeginEnd}[3]{}
\providecommand*{\EndOfBibitem}{}
\mciteSetBstSublistMode{f}
\mciteSetBstMaxWidthForm{subitem}
{(\emph{\alph{mcitesubitemcount}})}
\mciteSetBstSublistLabelBeginEnd{\mcitemaxwidthsubitemform\space}
{\relax}{\relax}

\bibitem[Landau and Lifshitz(1958)]{Landaubook}
L.~D. Landau and E.~M. Lifshitz, \emph{Statistical Physics}, Pergamon, Oxford,
  1st edn., 1958\relax
\mciteBstWouldAddEndPuncttrue
\mciteSetBstMidEndSepPunct{\mcitedefaultmidpunct}
{\mcitedefaultendpunct}{\mcitedefaultseppunct}\relax
\EndOfBibitem
\bibitem[Evans(1979)]{Evans}
R.~Evans, \emph{Adv.~Physics}, 1979, \textbf{28}, 143\relax
\mciteBstWouldAddEndPuncttrue
\mciteSetBstMidEndSepPunct{\mcitedefaultmidpunct}
{\mcitedefaultendpunct}{\mcitedefaultseppunct}\relax
\EndOfBibitem
\bibitem[de~Gennes(1971)]{pgdg71}
P.~de~Gennes, \emph{Mol. Cryst. Liq. Cryst.}, 1971, \textbf{12}, 191\relax
\mciteBstWouldAddEndPuncttrue
\mciteSetBstMidEndSepPunct{\mcitedefaultmidpunct}
{\mcitedefaultendpunct}{\mcitedefaultseppunct}\relax
\EndOfBibitem
\bibitem[de~Gennes(1973)]{pgdg73}
P.~de~Gennes, \emph{Mol. Cryst. Liq. Cryst.}, 1973, \textbf{21}, 49\relax
\mciteBstWouldAddEndPuncttrue
\mciteSetBstMidEndSepPunct{\mcitedefaultmidpunct}
{\mcitedefaultendpunct}{\mcitedefaultseppunct}\relax
\EndOfBibitem
\bibitem[L{\"o}wen \emph{et~al.}()L{\"o}wen, Beier, and Wagner]{Loewen1}
H.~L{\"o}wen, T.~Beier and H.~Wagner, \textit{Europhys.~Lett.}, 1989,
  \textbf{9}, 791; \textit{Z.~Phys.~B:~Condens.~Matter}, 1990, \textbf{79},
  109\relax
\mciteBstWouldAddEndPuncttrue
\mciteSetBstMidEndSepPunct{\mcitedefaultmidpunct}
{\mcitedefaultendpunct}{\mcitedefaultseppunct}\relax
\EndOfBibitem
\bibitem[Ohnesorge \emph{et~al.}(1991)Ohnesorge, L{\"o}wen, and
  Wagner]{Ohnesorge}
R.~Ohnesorge, H.~L{\"o}wen and H.~Wagner, \emph{Phys. Rev. A}, 1991,
  \textbf{43}, 2870\relax
\mciteBstWouldAddEndPuncttrue
\mciteSetBstMidEndSepPunct{\mcitedefaultmidpunct}
{\mcitedefaultendpunct}{\mcitedefaultseppunct}\relax
\EndOfBibitem
\bibitem[Lutsko(2006)]{Lutsko}
J.~F. Lutsko, \emph{Physica~A}, 2006, \textbf{366}, 229\relax
\mciteBstWouldAddEndPuncttrue
\mciteSetBstMidEndSepPunct{\mcitedefaultmidpunct}
{\mcitedefaultendpunct}{\mcitedefaultseppunct}\relax
\EndOfBibitem
\bibitem[Elder \emph{et~al.}(2002)Elder, Katakowski, Haataja, and Grant]{Mikko}
K.~R. Elder, M.~Katakowski, M.~Haataja and M.~Grant, \emph{Phys. Rev. Lett.},
  2002, \textbf{88}, 245701\relax
\mciteBstWouldAddEndPuncttrue
\mciteSetBstMidEndSepPunct{\mcitedefaultmidpunct}
{\mcitedefaultendpunct}{\mcitedefaultseppunct}\relax
\EndOfBibitem
\bibitem[Elder and Grant(2004)]{Elder_0}
K.~R. Elder and M.~Grant, \emph{Phys. Rev. E}, 2004, \textbf{70}, 051605\relax
\mciteBstWouldAddEndPuncttrue
\mciteSetBstMidEndSepPunct{\mcitedefaultmidpunct}
{\mcitedefaultendpunct}{\mcitedefaultseppunct}\relax
\EndOfBibitem
\bibitem[Emmerich(2009)]{Emmerich}
H.~Emmerich, \emph{J. Phys.: Condens. Matter}, 2009, \textbf{21}, 464103\relax
\mciteBstWouldAddEndPuncttrue
\mciteSetBstMidEndSepPunct{\mcitedefaultmidpunct}
{\mcitedefaultendpunct}{\mcitedefaultseppunct}\relax
\EndOfBibitem
\bibitem[Jaatinen and Ala-Nissila(2010)]{Jaatinnen_JPCM_2010}
A.~Jaatinen and T.~Ala-Nissila, \emph{J. Phys.: Condens. Matter}, 2010,
  \textbf{22}, 205402\relax
\mciteBstWouldAddEndPuncttrue
\mciteSetBstMidEndSepPunct{\mcitedefaultmidpunct}
{\mcitedefaultendpunct}{\mcitedefaultseppunct}\relax
\EndOfBibitem
\bibitem[Yu \emph{et~al.}(2009)Yu, Liu, and Voigt]{Axel}
Y.~M. Yu, B.~G. Liu and A.~Voigt, \emph{Phys. Rev. B}, 2009, \textbf{79},
  235317\relax
\mciteBstWouldAddEndPuncttrue
\mciteSetBstMidEndSepPunct{\mcitedefaultmidpunct}
{\mcitedefaultendpunct}{\mcitedefaultseppunct}\relax
\EndOfBibitem
\bibitem[Jaatinen \emph{et~al.}(2009)Jaatinen, Achim, Elder, and
  Ala-Nissila]{Tapio}
A.~Jaatinen, C.~V. Achim, K.~R. Elder and T.~Ala-Nissila, \emph{Phys. Rev. E},
  2009, \textbf{80}, 031602\relax
\mciteBstWouldAddEndPuncttrue
\mciteSetBstMidEndSepPunct{\mcitedefaultmidpunct}
{\mcitedefaultendpunct}{\mcitedefaultseppunct}\relax
\EndOfBibitem
\bibitem[Tegze \emph{et~al.}(2009)Tegze, Granasy, Toth, Podmaniczky, Jaatinen,
  Ala-Nissila, and Pusztai]{Laszlo}
G.~Tegze, L.~Granasy, G.~I. Toth, F.~Podmaniczky, A.~Jaatinen, T.~Ala-Nissila
  and T.~Pusztai, \emph{Phys. Rev. Lett.}, 2009, \textbf{103}, 035702\relax
\mciteBstWouldAddEndPuncttrue
\mciteSetBstMidEndSepPunct{\mcitedefaultmidpunct}
{\mcitedefaultendpunct}{\mcitedefaultseppunct}\relax
\EndOfBibitem
\bibitem[Mellenthin \emph{et~al.}(2008)Mellenthin, Karma, and Plapp]{Plapp}
J.~Mellenthin, A.~Karma and M.~Plapp, \emph{Phys. Rev. B}, 2008, \textbf{78},
  184110\relax
\mciteBstWouldAddEndPuncttrue
\mciteSetBstMidEndSepPunct{\mcitedefaultmidpunct}
{\mcitedefaultendpunct}{\mcitedefaultseppunct}\relax
\EndOfBibitem
\bibitem[McKenna \emph{et~al.}(2009)McKenna, Gururajan, and Voorhees]{Voorhees}
I.~M. McKenna, M.~P. Gururajan and P.~W. Voorhees, \emph{J. Mater. Sci.}, 2009,
  \textbf{44}, 2206--2217\relax
\mciteBstWouldAddEndPuncttrue
\mciteSetBstMidEndSepPunct{\mcitedefaultmidpunct}
{\mcitedefaultendpunct}{\mcitedefaultseppunct}\relax
\EndOfBibitem
\bibitem[Asaro and Tiller(1972)]{Asaro}
R.~J. Asaro and W.~A. Tiller, \emph{Metall. Trans.}, 1972, \textbf{3},
  1789\relax
\mciteBstWouldAddEndPuncttrue
\mciteSetBstMidEndSepPunct{\mcitedefaultmidpunct}
{\mcitedefaultendpunct}{\mcitedefaultseppunct}\relax
\EndOfBibitem
\bibitem[Grinfeld(1986)]{Grinfeld}
M.~A. Grinfeld, \emph{Sov. Phys. Dokl.}, 1986, \textbf{31}, 831\relax
\mciteBstWouldAddEndPuncttrue
\mciteSetBstMidEndSepPunct{\mcitedefaultmidpunct}
{\mcitedefaultendpunct}{\mcitedefaultseppunct}\relax
\EndOfBibitem
\bibitem[Huang and Elder(2008)]{Elder}
Z.~F. Huang and K.~R. Elder, \emph{Phys. Rev. Lett.}, 2008, \textbf{101},
  158701\relax
\mciteBstWouldAddEndPuncttrue
\mciteSetBstMidEndSepPunct{\mcitedefaultmidpunct}
{\mcitedefaultendpunct}{\mcitedefaultseppunct}\relax
\EndOfBibitem
\bibitem[Wu and Voorhees(2009)]{Voorhees2}
K.~A. Wu and P.~W. Voorhees, \emph{Phys. Rev. B}, 2009, \textbf{80},
  125408\relax
\mciteBstWouldAddEndPuncttrue
\mciteSetBstMidEndSepPunct{\mcitedefaultmidpunct}
{\mcitedefaultendpunct}{\mcitedefaultseppunct}\relax
\EndOfBibitem
\bibitem[Singh(1991)]{Singh:91}
Y.~Singh, \emph{Phys.~Reports}, 1991, \textbf{207}, 351\relax
\mciteBstWouldAddEndPuncttrue
\mciteSetBstMidEndSepPunct{\mcitedefaultmidpunct}
{\mcitedefaultendpunct}{\mcitedefaultseppunct}\relax
\EndOfBibitem
\bibitem[L{\"o}wen(1994)]{Loewen:94}
H.~L{\"o}wen, \emph{Phys.~Reports}, 1994, \textbf{237}, 249\relax
\mciteBstWouldAddEndPuncttrue
\mciteSetBstMidEndSepPunct{\mcitedefaultmidpunct}
{\mcitedefaultendpunct}{\mcitedefaultseppunct}\relax
\EndOfBibitem
\bibitem[van Teeffelen \emph{et~al.}(2009)van Teeffelen, L\"owen, Backofen, and
  Voigt]{Sven_09}
S.~van Teeffelen, H.~L\"owen, R.~Backofen and A.~Voigt, \emph{Phys. Rev. E},
  2009, \textbf{79}, 051404\relax
\mciteBstWouldAddEndPuncttrue
\mciteSetBstMidEndSepPunct{\mcitedefaultmidpunct}
{\mcitedefaultendpunct}{\mcitedefaultseppunct}\relax
\EndOfBibitem
\bibitem[Roth(2010)]{Roth_JPCM_2010}
R.~Roth, \emph{J. Phys.: Condens. Matter}, 2010, \textbf{22}, 063102\relax
\mciteBstWouldAddEndPuncttrue
\mciteSetBstMidEndSepPunct{\mcitedefaultmidpunct}
{\mcitedefaultendpunct}{\mcitedefaultseppunct}\relax
\EndOfBibitem
\bibitem[Ramakrishnan and Yussouff(1979)]{Ramakrishnan:79}
T.~V. Ramakrishnan and M.~Yussouff, \emph{Phys. Rev. B}, 1979, \textbf{19},
  2775\relax
\mciteBstWouldAddEndPuncttrue
\mciteSetBstMidEndSepPunct{\mcitedefaultmidpunct}
{\mcitedefaultendpunct}{\mcitedefaultseppunct}\relax
\EndOfBibitem
\bibitem[Rosenfeld \emph{et~al.}(1997)Rosenfeld, Schmidt, L{\"o}wen, and
  Tarazona]{FMT1}
Y.~Rosenfeld, M.~Schmidt, H.~L{\"o}wen and P.~Tarazona, \emph{Phys. Rev. E},
  1997, \textbf{55}, 4245\relax
\mciteBstWouldAddEndPuncttrue
\mciteSetBstMidEndSepPunct{\mcitedefaultmidpunct}
{\mcitedefaultendpunct}{\mcitedefaultseppunct}\relax
\EndOfBibitem
\bibitem[Roth \emph{et~al.}(2002)Roth, Evans, Lang, and Kahl]{White_bear}
R.~Roth, R.~Evans, A.~Lang and G.~Kahl, \emph{J. Phys.: Condens. Matter}, 2002,
  \textbf{14}, 12063--12078\relax
\mciteBstWouldAddEndPuncttrue
\mciteSetBstMidEndSepPunct{\mcitedefaultmidpunct}
{\mcitedefaultendpunct}{\mcitedefaultseppunct}\relax
\EndOfBibitem
\bibitem[Hansen-Goos and Mecke(2009)]{FMT2}
H.~Hansen-Goos and K.~Mecke, \emph{Phys. Rev. Lett.}, 2009, \textbf{102},
  018302\relax
\mciteBstWouldAddEndPuncttrue
\mciteSetBstMidEndSepPunct{\mcitedefaultmidpunct}
{\mcitedefaultendpunct}{\mcitedefaultseppunct}\relax
\EndOfBibitem
\bibitem[Elder \emph{et~al.}(2007)Elder, Provatas, Berry, Stefanovic, and
  Grant]{Provatas}
K.~R. Elder, N.~Provatas, J.~Berry, P.~Stefanovic and M.~Grant, \emph{Phys.
  Rev. B}, 2007, \textbf{75}, 064107\relax
\mciteBstWouldAddEndPuncttrue
\mciteSetBstMidEndSepPunct{\mcitedefaultmidpunct}
{\mcitedefaultendpunct}{\mcitedefaultseppunct}\relax
\EndOfBibitem
\bibitem[L{\"o}wen()]{Lowen_JPCM_2010}
H.~L{\"o}wen, \textit{J. Phys.: Condens. Matter}, 2010, In print\relax
\mciteBstWouldAddEndPuncttrue
\mciteSetBstMidEndSepPunct{\mcitedefaultmidpunct}
{\mcitedefaultendpunct}{\mcitedefaultseppunct}\relax
\EndOfBibitem
\bibitem[Bolhuis and Frenkel(1997)]{Bolhuis}
P.~Bolhuis and D.~Frenkel, \emph{J. Chem. Phys.}, 1997, \textbf{106},
  666--687\relax
\mciteBstWouldAddEndPuncttrue
\mciteSetBstMidEndSepPunct{\mcitedefaultmidpunct}
{\mcitedefaultendpunct}{\mcitedefaultseppunct}\relax
\EndOfBibitem
\bibitem[L{\"o}wen(1994)]{Lowen_HS}
H.~L{\"o}wen, \emph{Phys. Rev. E}, 1994, \textbf{50}, 1232\relax
\mciteBstWouldAddEndPuncttrue
\mciteSetBstMidEndSepPunct{\mcitedefaultmidpunct}
{\mcitedefaultendpunct}{\mcitedefaultseppunct}\relax
\EndOfBibitem
\bibitem[Frenkel \emph{et~al.}(1984)Frenkel, Mulder, and McTague]{Mulder}
D.~Frenkel, B.~M. Mulder and J.~P. McTague, \emph{Phys. Rev. Lett.}, 1984,
  \textbf{52}, 287--290\relax
\mciteBstWouldAddEndPuncttrue
\mciteSetBstMidEndSepPunct{\mcitedefaultmidpunct}
{\mcitedefaultendpunct}{\mcitedefaultseppunct}\relax
\EndOfBibitem
\bibitem[L{\"o}wen(1994)]{Yukawa_Lowen_1}
H.~L{\"o}wen, \emph{Phys. Rev. Lett.}, 1994, \textbf{72}, 424\relax
\mciteBstWouldAddEndPuncttrue
\mciteSetBstMidEndSepPunct{\mcitedefaultmidpunct}
{\mcitedefaultendpunct}{\mcitedefaultseppunct}\relax
\EndOfBibitem
\bibitem[L{\"o}wen(1994)]{Yukawa_Lowen_2}
H.~L{\"o}wen, \emph{J. Chem. Phys.}, 1994, \textbf{100}, 6738\relax
\mciteBstWouldAddEndPuncttrue
\mciteSetBstMidEndSepPunct{\mcitedefaultmidpunct}
{\mcitedefaultendpunct}{\mcitedefaultseppunct}\relax
\EndOfBibitem
\bibitem[Kirchhoff \emph{et~al.}(1996)Kirchhoff, L{\"o}wen, and
  Klein]{Yukawa_Lowen_3}
T.~Kirchhoff, H.~L{\"o}wen and R.~Klein, \emph{Phys. Rev. E}, 1996,
  \textbf{53}, 5011\relax
\mciteBstWouldAddEndPuncttrue
\mciteSetBstMidEndSepPunct{\mcitedefaultmidpunct}
{\mcitedefaultendpunct}{\mcitedefaultseppunct}\relax
\EndOfBibitem
\bibitem[Cleaver \emph{et~al.}(1996)Cleaver, Care, Allen, and
  Neal]{Gay_Berne_1}
D.~J. Cleaver, C.~M. Care, M.~P. Allen and M.~P. Neal, \emph{Phys. Rev. E},
  1996, \textbf{54}, 559--567\relax
\mciteBstWouldAddEndPuncttrue
\mciteSetBstMidEndSepPunct{\mcitedefaultmidpunct}
{\mcitedefaultendpunct}{\mcitedefaultseppunct}\relax
\EndOfBibitem
\bibitem[Fukunaga \emph{et~al.}(2004)Fukunaga, Takimoto, and Doi]{Gay_Berne_2}
H.~Fukunaga, J.~Takimoto and M.~Doi, \emph{J. Chem. Phys.}, 2004, \textbf{120},
  7792--7800\relax
\mciteBstWouldAddEndPuncttrue
\mciteSetBstMidEndSepPunct{\mcitedefaultmidpunct}
{\mcitedefaultendpunct}{\mcitedefaultseppunct}\relax
\EndOfBibitem
\bibitem[Muccioli and Zannoni(2006)]{Gay_Berne_3}
L.~Muccioli and C.~Zannoni, \emph{Chem. Phys. Lett.}, 2006, \textbf{423},
  1--6\relax
\mciteBstWouldAddEndPuncttrue
\mciteSetBstMidEndSepPunct{\mcitedefaultmidpunct}
{\mcitedefaultendpunct}{\mcitedefaultseppunct}\relax
\EndOfBibitem
\bibitem[Frenkel(1991)]{Frenkel}
D.~Frenkel, \emph{Liquids, Freezing and the Glass Transition}, Les Houches
  Summer Schools of Theoretical Physics, Amsterdam, 1991\relax
\mciteBstWouldAddEndPuncttrue
\mciteSetBstMidEndSepPunct{\mcitedefaultmidpunct}
{\mcitedefaultendpunct}{\mcitedefaultseppunct}\relax
\EndOfBibitem
\bibitem[Rex \emph{et~al.}(2007)Rex, Wensink, and L\"owen]{Rex_08}
M.~Rex, H.~H. Wensink and H.~L\"owen, \emph{Phys. Rev. E}, 2007, \textbf{76},
  021403\relax
\mciteBstWouldAddEndPuncttrue
\mciteSetBstMidEndSepPunct{\mcitedefaultmidpunct}
{\mcitedefaultendpunct}{\mcitedefaultseppunct}\relax
\EndOfBibitem
\bibitem[Poniewierski and Holyst(1988)]{Holyst}
A.~Poniewierski and R.~Holyst, \emph{Phys. Rev. Lett.}, 1988, \textbf{61},
  2461--2464\relax
\mciteBstWouldAddEndPuncttrue
\mciteSetBstMidEndSepPunct{\mcitedefaultmidpunct}
{\mcitedefaultendpunct}{\mcitedefaultseppunct}\relax
\EndOfBibitem
\bibitem[Graf and L{\"o}wen(1999)]{Graf}
H.~Graf and H.~L{\"o}wen, \emph{J. Phys.: Condens. Matter}, 1999, \textbf{11},
  1435\relax
\mciteBstWouldAddEndPuncttrue
\mciteSetBstMidEndSepPunct{\mcitedefaultmidpunct}
{\mcitedefaultendpunct}{\mcitedefaultseppunct}\relax
\EndOfBibitem
\bibitem[foo()]{footnote_Referee_B}
As an equivalent description, a position-dependent nematic tensor could be used
  in the density parametrization instead of using the fields
  $\psi_{2}(\vec{R})$ and $\hat{u}_{0}(\vec{R})$.\relax
\mciteBstWouldAddEndPunctfalse
\mciteSetBstMidEndSepPunct{\mcitedefaultmidpunct}
{}{\mcitedefaultseppunct}\relax
\EndOfBibitem
\bibitem[Graf and L{\"o}wen(1998)]{GrafLoewen1998}
H.~Graf and H.~L{\"o}wen, \emph{Phys. Rev. E}, 1998, \textbf{57},
  5744--575\relax
\mciteBstWouldAddEndPuncttrue
\mciteSetBstMidEndSepPunct{\mcitedefaultmidpunct}
{\mcitedefaultendpunct}{\mcitedefaultseppunct}\relax
\EndOfBibitem
\bibitem[de~Gennes and Prost(1993)]{pgdgbook}
P.~de~Gennes and J.~Prost, \emph{The Physics of Liquid Crystals}, Clarendon
  Press, Oxford, 1993\relax
\mciteBstWouldAddEndPuncttrue
\mciteSetBstMidEndSepPunct{\mcitedefaultmidpunct}
{\mcitedefaultendpunct}{\mcitedefaultseppunct}\relax
\EndOfBibitem
\bibitem[Mukjherjee \emph{et~al.}(2001)Mukjherjee, Pleiner, and Brand]{mpb1}
P.~Mukjherjee, H.~Pleiner and H.~Brand, \emph{Eur. Phys. J E}, 2001,
  \textbf{4}, 293\relax
\mciteBstWouldAddEndPuncttrue
\mciteSetBstMidEndSepPunct{\mcitedefaultmidpunct}
{\mcitedefaultendpunct}{\mcitedefaultseppunct}\relax
\EndOfBibitem
\bibitem[Chaikin and Lubensky(1995)]{cl}
P.~Chaikin and T.~Lubensky, \emph{Principles of Condensed Matter Physics},
  Cambridge University Press, Cambridge, 1995\relax
\mciteBstWouldAddEndPuncttrue
\mciteSetBstMidEndSepPunct{\mcitedefaultmidpunct}
{\mcitedefaultendpunct}{\mcitedefaultseppunct}\relax
\EndOfBibitem
\bibitem[de~Gennes(1969)]{pgdg69}
P.~de~Gennes, \emph{J. Phys. Coll.}, 1969, \textbf{30 - C4}, 65\relax
\mciteBstWouldAddEndPuncttrue
\mciteSetBstMidEndSepPunct{\mcitedefaultmidpunct}
{\mcitedefaultendpunct}{\mcitedefaultseppunct}\relax
\EndOfBibitem
\bibitem[Frank(1958)]{frank}
F.~Frank, \emph{Discuss. Faraday Soc.}, 1958, \textbf{25}, 19\relax
\mciteBstWouldAddEndPuncttrue
\mciteSetBstMidEndSepPunct{\mcitedefaultmidpunct}
{\mcitedefaultendpunct}{\mcitedefaultseppunct}\relax
\EndOfBibitem
\bibitem[Brand and Kawasaki(1986)]{bk86}
H.~Brand and K.~Kawasaki, \emph{J. Phys. C}, 1986, \textbf{19}, 937\relax
\mciteBstWouldAddEndPuncttrue
\mciteSetBstMidEndSepPunct{\mcitedefaultmidpunct}
{\mcitedefaultendpunct}{\mcitedefaultseppunct}\relax
\EndOfBibitem
\bibitem[Brand and Pleiner(1987)]{bp87}
H.~Brand and H.~Pleiner, \emph{Phys. Rev. A}, 1987, \textbf{35}, 3122\relax
\mciteBstWouldAddEndPuncttrue
\mciteSetBstMidEndSepPunct{\mcitedefaultmidpunct}
{\mcitedefaultendpunct}{\mcitedefaultseppunct}\relax
\EndOfBibitem
\bibitem[Cross(1975)]{mcc75}
M.~Cross, \emph{J. Low Temp. Phys.}, 1975, \textbf{21}, 525\relax
\mciteBstWouldAddEndPuncttrue
\mciteSetBstMidEndSepPunct{\mcitedefaultmidpunct}
{\mcitedefaultendpunct}{\mcitedefaultseppunct}\relax
\EndOfBibitem
\bibitem[Pleiner and Brand(1980)]{pb80}
H.~Pleiner and H.~Brand, \emph{J. Phys. (Paris) Lett.}, 1980, \textbf{41},
  491\relax
\mciteBstWouldAddEndPuncttrue
\mciteSetBstMidEndSepPunct{\mcitedefaultmidpunct}
{\mcitedefaultendpunct}{\mcitedefaultseppunct}\relax
\EndOfBibitem
\bibitem[Pleiner and Brand(1995)]{pb95}
H.~Pleiner and H.~R. Brand, in \emph{Pattern Formation in Liquid Crystals}, ed.
  A.~Buka and L.~Kramer, Springer, New York, 1995, ch. Hydrodynamics and
  Electrohydrodynamics of Liquid Crystals, p.~15\relax
\mciteBstWouldAddEndPuncttrue
\mciteSetBstMidEndSepPunct{\mcitedefaultmidpunct}
{\mcitedefaultendpunct}{\mcitedefaultseppunct}\relax
\EndOfBibitem
\bibitem[McDonald \emph{et~al.}(2001)McDonald, Allen, and Schmid]{Schmid}
A.~J. McDonald, M.~P. Allen and F.~Schmid, \emph{Phys. Rev. E}, 2001,
  \textbf{63}, 010701\relax
\mciteBstWouldAddEndPuncttrue
\mciteSetBstMidEndSepPunct{\mcitedefaultmidpunct}
{\mcitedefaultendpunct}{\mcitedefaultseppunct}\relax
\EndOfBibitem
\bibitem[{van der Beek} \emph{et~al.}(2006){van der Beek}, Reich, {van der
  Schoot}, Dijkstra, Schilling, {L. V. Vink}, Schmidt, {van Roij}, and
  Lekkerkerker]{Beek}
D.~{van der Beek}, H.~Reich, P.~{van der Schoot}, M.~Dijkstra, T.~Schilling,
  R.~{L. V. Vink}, M.~Schmidt, R.~{van Roij} and H.~Lekkerkerker, \emph{Phys.
  Rev. Lett.}, 2006, \textbf{97}, 087801\relax
\mciteBstWouldAddEndPuncttrue
\mciteSetBstMidEndSepPunct{\mcitedefaultmidpunct}
{\mcitedefaultendpunct}{\mcitedefaultseppunct}\relax
\EndOfBibitem
\bibitem[Bier \emph{et~al.}(2005)Bier, Harnau, and Dietrich]{Harnau}
M.~Bier, L.~Harnau and S.~Dietrich, \emph{J. Chem. Phys.}, 2005, \textbf{123},
  114906\relax
\mciteBstWouldAddEndPuncttrue
\mciteSetBstMidEndSepPunct{\mcitedefaultmidpunct}
{\mcitedefaultendpunct}{\mcitedefaultseppunct}\relax
\EndOfBibitem
\bibitem[Ohnesorge \emph{et~al.}(1993)Ohnesorge, L{\"o}wen, and
  Wagner]{EPL_1993_Ohnesorge}
R.~Ohnesorge, H.~L{\"o}wen and H.~Wagner, \emph{Europhys. Letters}, 1993,
  \textbf{22}, 245\relax
\mciteBstWouldAddEndPuncttrue
\mciteSetBstMidEndSepPunct{\mcitedefaultmidpunct}
{\mcitedefaultendpunct}{\mcitedefaultseppunct}\relax
\EndOfBibitem
\bibitem[van Roij \emph{et~al.}(1995)van Roij, Bolhuis, Mulder, and
  Frenkel]{vanRoij}
R.~van Roij, P.~Bolhuis, B.~Mulder and D.~Frenkel, \emph{Phys. Rev. E}, 1995,
  \textbf{52}, R1277--R1280\relax
\mciteBstWouldAddEndPuncttrue
\mciteSetBstMidEndSepPunct{\mcitedefaultmidpunct}
{\mcitedefaultendpunct}{\mcitedefaultseppunct}\relax
\EndOfBibitem
\bibitem[Marconi and Tarazona(1999)]{Marconi}
U.~M.~B. Marconi and P.~Tarazona, \emph{J. Chem. Phys.}, 1999, \textbf{110},
  8032\relax
\mciteBstWouldAddEndPuncttrue
\mciteSetBstMidEndSepPunct{\mcitedefaultmidpunct}
{\mcitedefaultendpunct}{\mcitedefaultseppunct}\relax
\EndOfBibitem
\bibitem[Archer and Evans(2004)]{Archer_Evans}
A.~J. Archer and R.~Evans, \emph{J. Chem. Phys.}, 2004, \textbf{121},
  4246\relax
\mciteBstWouldAddEndPuncttrue
\mciteSetBstMidEndSepPunct{\mcitedefaultmidpunct}
{\mcitedefaultendpunct}{\mcitedefaultseppunct}\relax
\EndOfBibitem
\bibitem[Espanol and L\"owen(2009)]{Pep}
P.~Espanol and H.~L\"owen, \emph{J. Chem. Phys.}, 2009, \textbf{131},
  244101\relax
\mciteBstWouldAddEndPuncttrue
\mciteSetBstMidEndSepPunct{\mcitedefaultmidpunct}
{\mcitedefaultendpunct}{\mcitedefaultseppunct}\relax
\EndOfBibitem
\bibitem[Liu and Muthukumar(1997)]{Muthukumar}
C.~Liu and M.~Muthukumar, \emph{J. Chem. Phys.}, 1997, \textbf{106}, 7822\relax
\mciteBstWouldAddEndPuncttrue
\mciteSetBstMidEndSepPunct{\mcitedefaultmidpunct}
{\mcitedefaultendpunct}{\mcitedefaultseppunct}\relax
\EndOfBibitem
\bibitem[Bechhoefer \emph{et~al.}(1991)Bechhoefer, L{\"o}wen, and
  Tuckerman]{Bechhoefer}
J.~Bechhoefer, H.~L{\"o}wen and L.~S. Tuckerman, \emph{Phys. Rev. Lett.}, 1991,
  \textbf{67}, 1266\relax
\mciteBstWouldAddEndPuncttrue
\mciteSetBstMidEndSepPunct{\mcitedefaultmidpunct}
{\mcitedefaultendpunct}{\mcitedefaultseppunct}\relax
\EndOfBibitem
\bibitem[Renner \emph{et~al.}(1995)Renner, L{\"o}wen, and
  Barrat]{Renner_Barrat}
C.~Renner, H.~L{\"o}wen and J.-L. Barrat, \emph{Phys. Rev. E}, 1995,
  \textbf{5}, 5091\relax
\mciteBstWouldAddEndPuncttrue
\mciteSetBstMidEndSepPunct{\mcitedefaultmidpunct}
{\mcitedefaultendpunct}{\mcitedefaultseppunct}\relax
\EndOfBibitem
\bibitem[Toner \emph{et~al.}(2005)Toner, Tu, and Ramaswamy]{Ramaswamy}
J.~Toner, Y.~Tu and S.~Ramaswamy, \emph{Annals of Physics}, 2005, \textbf{318},
  170\relax
\mciteBstWouldAddEndPuncttrue
\mciteSetBstMidEndSepPunct{\mcitedefaultmidpunct}
{\mcitedefaultendpunct}{\mcitedefaultseppunct}\relax
\EndOfBibitem
\bibitem[Peruani \emph{et~al.}(2006)Peruani, Deutsch, and B\"ar]{Markus_Baer}
F.~Peruani, A.~Deutsch and M.~B\"ar, \emph{Phys. Rev. E}, 2006, \textbf{74},
  030904(R)\relax
\mciteBstWouldAddEndPuncttrue
\mciteSetBstMidEndSepPunct{\mcitedefaultmidpunct}
{\mcitedefaultendpunct}{\mcitedefaultseppunct}\relax
\EndOfBibitem
\bibitem[Wensink and L{\"o}wen(2008)]{Wensink_08}
H.~H. Wensink and H.~L{\"o}wen, \emph{Phys. Rev. E}, 2008, \textbf{78},
  031409\relax
\mciteBstWouldAddEndPuncttrue
\mciteSetBstMidEndSepPunct{\mcitedefaultmidpunct}
{\mcitedefaultendpunct}{\mcitedefaultseppunct}\relax
\EndOfBibitem
\end{mcitethebibliography}
\bibliographystyle{rsc}
\end{document}